\documentclass[a4paper, amsfonts, amssymb, amsmath, reprint, showkeys, nofootinbib, twoside,superscriptaddress, aps, prm]{revtex4-2}
\usepackage[english]{babel}
\usepackage[utf8]{inputenc}
\usepackage[colorinlistoftodos, color=green!40, prependcaption]{todonotes}
\usepackage{amsthm}
\usepackage{mathtools}
\usepackage{physics}
\usepackage{xcolor}
\usepackage{graphicx}
\usepackage{mathtools, bm, comment}
\usepackage[left=23mm,right=13mm,top=35mm,columnsep=15pt]{geometry} 
\usepackage{adjustbox}
\usepackage{placeins}
\usepackage[T1]{fontenc}
\usepackage{csquotes}
\usepackage[caption=false]{subfig}
\usepackage{subfig}
\usepackage{multirow}
\usepackage{bigints}
\usepackage[pdftex, pdftitle={Article}, pdfauthor={Author}]{hyperref}
\begin{document}
	
	\title{The influence of hydrogen on the electronic structure in transition metallic glasses}
	
	\author{Johan Bylin}
	\affiliation{Division of Materials Physics, Department of Physics and Astronomy, Uppsala University, Box 516, SE-75121, Uppsala, Sweden}
	\author{Rebecka Lindblad}
	\affiliation{Division of X-ray Photon Science, Department of Physics and Astronomy, Uppsala University, Box 516, SE-75121, Uppsala, Sweden}
	\author{Lennart Spode}
	\affiliation{Division of Materials Physics, Department of Physics and Astronomy, Uppsala University, Box 516, SE-75121, Uppsala, Sweden}
	\author{Ralph H. Scheicher}
	\affiliation{Division of Materials Theory, Department of Physics and Astronomy, Uppsala University, Box 516, SE-75121, Uppsala, Sweden}
	\author{Gunnar K. P\'alsson}
	\email[Correspondence email address: ]{gunnar.palsson@physics.uu.se}
	\affiliation{Division of Materials Physics, Department of Physics and Astronomy, Uppsala University, Box 516, SE-75121, Uppsala, Sweden}
	\date{\today}
	
	\begin{abstract}
		We investigate the influence of hydrogen on the electronic structure of a binary transition metallic glass of V$_{80}$Zr$_{20}$. We examine the hybridization between the hydrogen and metal atoms with the aid of hard x-ray photoelectron spectroscopy. Combined with \emph{ab initio} density functional theory, we are able to show and predict the formation of $s$-$d$ hybridized energy states. With optical transmission and resistivity measurements, we investigate the emergent electronic properties formed out of those altered energy states, and together with the theoretical calculations of the frequency-dependent conductivity tensor, we qualitatively support the observed strong wavelength-dependency of the hydrogen-induced changes on the optical absorption and a positive parabolic change in resistivity with hydrogen concentration. 
	\end{abstract}
	

	\maketitle
	
	\section{Introduction} \label{sec:introduction}
	The electronic structure of materials underpins many aspects of condensed matter physics. The properties formed from the ground state and allowed electron transitions are responsible for the chemical, conductive, magnetic, optical, and thermodynamic behaviours of materials. The theoretical framework describing the phenomena is for many crystalline materials well established. Bloch waves and their respective energy eigenvalues in the form of electron bands are the fundamental building blocks for the description of the electronic properties of crystals. But, for disordered materials, such as metallic glasses, the translational symmetry is absent, and the concept of a bandstructure is rendered difficult to conceptualize~\cite{10.1063/1.457564}. Nevertheless, the emerging properties persist and can be measured, such as band gaps~\cite{K_Hulls_1972, COHEN1970391}, Hall effect~\cite{McGuire1980}, magneto-optic coupling~\cite{Hubert2008, zvezdin1997modern}, and optical conductivity~\cite{10.1063/1.3529465, 10.1063/1.2821376, https://doi.org/10.1002/adma.200602560}. 
	
	Many aspects of the structure of metallic glasses, such as the stability and the anomalous temperature dependence of resistivity, have been linked to the shape of the density of states at the Fermi level~\cite{PhysRevLett.35.380, PhysRevB.16.1694, doi:10.1080/14786436508211931}. Hydrogen has been known to improve the stability of metallic glasses, which can be intuitively understood in terms of the mixing of enthalpies in the semi-empirical Miedmea rules~\cite{MIEDEMA19801, INOUE2000279}. Yet, it raises the question: how do the density of states and the electronic properties, such as the optical conductivity, change with hydrogen concentration, and how can these changes be understood at the level of the electronic structure?
	
	Hydrogen dissolved into metals can be regarded as the simplest impurity or alloying problem. Its point-defect qualities not only elicit elastic strain of the host atomic structure, akin to centers of dilation~\cite{10.1080/00018737400101401, 10.1007/bfb0045966, 10.1103/physrevb.96.224103}, but the addition of a single proton and an electron can be regarded as one of the simplest elemental perturbations to the electronic structure of the metal~\cite{10.1524/zpch.1979.117.117.089, MANCHESTER19761, 10.1002/bbpc.19720760809, PhysRevB.20.446, PhysRevB.20.3543, PhysRevB.17.3518}. The effect has bearings on a number of related features of the metal-hydrogen system, such as its electron transport~\cite{doi:10.1143/JPSJ.54.3406, PRYDE19711333, 10.1063/1.3529465, SzafrańskiFilipek19861525} and optical properties~\cite{Victoria2016, https://doi.org/10.1002/adfm.202010483, 10.1063/1.3529465, 10.1063/1.2821376, Huiberts1996}. The electronic structure alterations have been observed in both crystalline~\cite{TANAKA1979173, IVANOVSKY19901, HUANG20043499, https://doi.org/10.1002/cphc.201801185, PhysRevB.58.5230} and amorphous~\cite{TANAKA1982317, PhysRevB.37.6215, SMFries_1985, PhysRevB.52.7151} transition metal hydrogen systems, whereby the formation of a $s$-$d$ hybridized band around 7 eV below the Fermi level has been attributed to the combination of the $s$ electrons of the hydrogen with the valence $d$ electrons of the metal.
	
	Switendick~\cite{10.1524/zpch.1979.117.117.089} was one of the first to perform band structure calculations of metal hydrides in order to understand this phenomenon, whereby he ascribed the stability of the hydride formation to the lowering and emergence of the $s$-$d$ hybridized band. Later on, N\o{}rskov~\cite{PhysRevB.20.446} introduced the hydrogen atom as a perturbation to the electronic structure of the metal and found an electronic repulsive hydrogen-hydrogen interaction between the hydrogen atoms embedded in the metal. Griessen~\cite{10.1103/physrevb.27.7575} and Richards~\cite{10.1103/physrevb.30.5183} stressed its importance and showed that it is a part of the mechanism that partly governs the phase stability of hydrogen in metals. However, unlike crystals, where the system can be decomposed into irreducible symmetry representations making the calculations tractable, amorphous materials on the other hand cannot, and remains a considerable challenge. While there have been attempts to model the electronic structure and properties of liquids and amorphous materials~\cite{doi:10.1080/14786437208220339, 10.1063/1.457564, S_Sinha_1989, J_Hafner_1988}, fully first-principles calculations have been historically unfeasible to conduct and it was not until the recent advent of supercomputers that representative supercells could be created~\cite{Holmstr} to accurately calculate the electronic properties of these highly disordered metal systems~\cite{PhysRevB.42.10887, PhysRevB.84.054203, Sun2020, Chen2018, SHA201516, ZhengGuang-Ping2012Adft}. 
	
	To the present day, no first-principles study has to our knowledge been conducted for hydrogen in metallic glasses. Combining state-of-the-art measurements, such as hard x-ray photoemission spectroscopy, together with \emph{ab initio} calculations we therefore seek to gain insight on the topic regarding the electronic structure of these disordered amorphous metal-hydrogen systems. 
	
	The electronic properties emergent from the transitions of the electronic bands close to and around the Fermi level can be captured concisely in the frequency-dependent optical conductivity $\bm{\sigma}(\omega)$. In fact, both the complex refractive index, $N = n + ik$, and direct current (DC) resistivity, $\rho_{\text{DC}}$, can be extracted from $\bm{\sigma}$; both of which are readily accessible experimentally.
	
	Here we utilize hard x-ray photoelectron spectroscopy (HAXPES) together with \emph{in situ} optical transmission and electrical resistivity techniques to study the effects of combining hydrogen with amorphous metals on the electronic structure of the metal-hydrogen system. For this purpose, we have opted to utilize rapidly quenched metallic thin films of amorphous vanadium zirconium V$_{80}$Zr$_{20}$, not only because of its practical applications~\cite{10.1103/physrevb.66.094109, 10.1016/0022-5088(84)90457-0, 10.1016/j.jallcom.2006.09.065}, but also because it readily absorbs hydrogen and is mechanically and thermally stable~\cite{eickert_formation_1992, KAPLAN2022101496, PhysRevB.106.104110}.

	
	\section{Experimental and Theoretical Details} \label{sec:ExperimentalDetails}
	\subsection{Sample preparation}
	Two sets of sample configurations of V$_{80}$Zr$_{20}$ thin films were grown on amorphous SiO$_2$ substrates using combinatorial DC magnetron sputtering. The first set was capped with Pd, while the other set, consisting of two simultaneously grown samples, was deposited with $6$ nm of Al$_2$O$_3$ as capping layer. Both sets of samples were synthesized under ultra-high-vacuum conditions with a base pressure below $3 \cdot10^{-10}$ Torr. As inert sputtering gas, $2.5$ mTorr of 6N argon was used, which was further filtered through a Nupure Omni 40 PF purifying filter. The substrates were pre-annealed at $623$ K for $30$ min in ultra-high vacuum to evaporate off water and other possible impurities on the surface. Before co-sputtering vanadium and zirconium, the temperature was allowed to settle to room temperature to maximize the quenching rate and thus promote maximally amorphous growth. The deposition rates of both the vanadium and zirconium targets were calibrated as a function of magnetron power to achieve accurate and desired amorphous composition. The composition was verified within $1$ at\% using Rutherford backscattering spectrometry. Complementary x-ray reflectivity and x-ray diffraction measurements were used to verify low interlayer roughness and that the samples were x-ray amorphous. The thickness of the film and capping layers were fitted using a slab model via the fitting program GenX~\cite{10.1107/s0021889807045086, 10.1107/s0021889811041446}. The thicknesses of the Pd and V$_{80}$Zr$_{20}$ layers were found to be $5.7(1)$ nm and $52.3(4)$ nm respectively. 
	
	\subsection{Hydrogen loading for HAXPES}
	An ultra-high-vacuum sample environment with a base pressure of $7.5\cdot 10^{-9}$ Torr was used to load one of the V$_{80}$Zr$_{20}$ samples capped with Al$_2$O$_3$ with hydrogen. The sample was exposed to hydrogen gas from a commercial hydrogen gas bottle of 6N purity. The sample was left in a 375 Torr hydrogen-pressure atmosphere at 423 K for a total of five days to equilibrate the hydrogen concentration in the sample before the sample was transferred for the HAXPES measurements.
	
	\subsection{HAXPES measurement}
	Hard x-ray Photoelectron Spectroscopy (HAXPES) was used to study the core and valence electronic structure of pristine and hydrogen-loaded V$_{80}$Zr$_{20}$ through the Al$_2$O$_3$ capping layer. The measurements were performed at the Kai Siegbahn laboratory at Uppsala University using a Scienta Omicron HAXPES lab equipped with a liquid Ga K$\alpha_1$ ($9250$ eV) x-ray source and an EW4000 electron energy analyser. The photon bandwidth of the Ga K$\alpha_1$ radiation is $0.5$ eV. The valence band was measured with an analyser slit of $0.5$ mm and a pass energy of $300$ eV giving an analyser resolution of $0.38$ eV. Combining the photon bandwidth with the analyser resolution gives a total experimental resolution of $0.62$ eV. The pressure in the chamber was $1\cdot 10^{-9}$ Torr during measurements. We note that the Al$_2$O$_3$ capping layer contributes to the valence spectrum in the binding energy region 5 eV to 15 eV and that this contribution is equal for the two samples.
	
	\subsection{Optical transmission and resistivity measurements}
	
	An ultra-high-vacuum chamber, capable of \emph{in situ} loading of hydrogen in a wide temperature and pressure range (see Ref.~\cite{10.1063/1.3529465} for schematics), was used to measure the changes in the optical transmission and the DC resistivity. Two light sources, 625 nm and 455 nm wavelengths (Thorlabs models M625F2 and M455F3, bandwidth 25 nm), were simultaneously used, each aligned to two separate silicon photodetectors (model PDA10A2 and APD430A2). The two detectors were connected to SR830 Stanford research lock-ins, each frequency-modulated with two separate relative prime lock-in frequencies to eliminate stray light contamination. The setup was capable of measuring electrical resistance via the four-probe method using current reversal of $0.1$ mA, set by a Keithley 2400 sourcemeter, which in trig-link mode with a Keithley 2182A nanovoltmeter measured the potential difference across the sample. Due to the Pd top layer, the sample stack was modeled as a parallel resistor, and with temperature-dependent tabulated values from NIST~\cite{10.1063/1.555614} along with the measured resistivity from a reference single layer 6 nm Pd calibration sample and the thickness of the individual layers, the resistivity changes of V$_{80}$Zr$_{20}$ with hydrogen could be extracted. The thickness expansion of the V$_{80}$Zr$_{20}$ as a function of hydrogen concentration, as measured in Ref.~\cite{PhysRevB.106.104110}, was taken into account when converting the sheet resistance to resistivity. 
	
	The gas pressures within the chamber were measured using a 10 Torr and a 1000 Torr Inficon CDG045 gauge and the pressure was controlled by a MKS type 244 flow control unit that adjusted the inlet gas flow via a motor controlled valve. To allow for desorption and excess gas pressure to escape, a pneumatic outlet valve kept the hydrogen pressure within a desired set-point pressure environment. The temperature of the sample was probed by a type-K thermocouple, in direct contact with the Pd top layer, and maintained within 0.2 K of its target temperature set-point via Hemi heating bands attached to the exterior of the chamber and a Eurotherm 2408 with fine-tuned PID values for different pressure points. Before the sample was exposed to hydrogen, the system was heated to 423 K in vacuum to evaporate off water and other volatile contaminants. The base pressure was below $5\cdot10^{-9}$ Torr before hydrogen was introduced.  
	
	Four absorption-desorption cycles were conducted at three different temperatures, 398 K, 423 K, and 448 K. For each of the isotherms, the pressure was dosed in steps and maintained at those pressures until temperature, resistance, and optical transmission were stable and the hydrogen concentration in the sample had reached equilibrium. The equilibrium condition, besides being within the target temperature and pressure set-point, was deemed to have been satisfied when the changes in resistance, sensitive to microstructural changes in the material, were stable within 0.1 m$\Omega /$minute. Once equilibrium was reached, an average data point over the last half hour of data points of pressure, temperature, resistance, and optical transmission was taken before the chamber was dosed with hydrogen to a new target pressure set-point. In total, the pressure range was set between $1\cdot10^{-3}$ Torr to $2\cdot 10^{2}$ Torr in 20 logarithmically equidistant steps.
	
	The analysis of the optical transmission was performed by employing Beer-Lamberts law, $I = I_0e^{-\alpha t}$, in combination with the fact that the absorption coefficient can be expressed in terms of the imaginary component of the refractive index, $\alpha = \frac{4\pi}{\lambda}k$, to relate the relative changes in transmission as,
	
	\begin{equation}\label{eq:meas_delta_k}
		\frac{\lambda}{4\pi t_0} \ln \left( \frac{ I_{\text{Tr}}( c_{\text{H}} ) }{I_{\text{Tr}}( 0 )} \right) = k( c_{\text{H}} )\frac{ t( c_{\text{H}} )} {t_0} - k_0.
	\end{equation}
	
	\noindent
	Here, $\lambda$ is the wavelength of the transmitted light, $I_{\text{Tr}}$ is the transmitted intensity, $c_{\text{H}}$ the hydrogen concentration in units of hydrogen per metal atoms H/M, $k_0$ is the hydrogen-free extinction coefficient and $t( c_{\text{H}} ) $ and $t_0$ is the thickness of the metal film with and without hydrogen respectively. As the hydrogen concentration in the palladium, within the measured temperature and hydrogen pressure ranges, was negligible~\cite{10.1007.bf02667685, PhysRevB.106.104110}, the relative changes represented in eq.~\ref{eq:meas_delta_k} were attributed to changes in the thickness and absorption coefficient of the metallic glass layer. Relating the changes in resistivity and optical transmission to the hydrogen concentration was done by converting the measured pressure to an effective hydrogen concentration via the explicit pressure and hydrogen concentration relationships measured with neutrons in Ref.~\cite{PhysRevB.106.104110} for each of the three temperature isotherms.
	
	The experimental transmission data, besides the first absorption cycling at $423$ K, has been shifted down to the magnitude of the transmission of the first cycle. This offset is attributed to thermal- and strain-induced relaxations that can be found in many amorphous materials~\cite{PhysRevB.30.5516, KAPLAN2022101496}. The effect has been ascribed to the filling of the vacant free volumes in the amorphous structure due to atomic reconfiguration. This causes a predictable and time-dependent drift in the resistivity, and here also in optical transmission, that occur together with the changes introduced by hydrogen. However, we argue that the relaxations are of benign character in the case of thin film V$_{80}$Zr$_{20}$ with respect to hydrogen cycling as several contributing results highlight the reversibility of the material with respect to hydrogen uptake and release. Firstly, as can be seen in Fig.~\ref{fig:Tr_vs_C}, all the transmission changes after the initial cycle exhibit practically identical behavior, regardless of the measured temperature, which is a significant result that emphasizes the electronic repeatability of the material when it comes to its hydrogen-dependent interaction with light.  Secondly, the changes in resistivity shown in Fig.~\ref{fig:R_vs_C}(a), treated and compensated for the drift in the same way, exhibit similar characteristics as the optical transmission, in which besides the first loading cycle, all other resistivity curves above $0.4$ H/M overlap,  see Fig.~\ref{fig:R_vs_C}. Thirdly, a linearly reversible volume expansion of thin film V$_{80}$Zr$_{20}$ was found in Ref.~\cite{PhysRevB.106.104110}, also irrespective of the measured temperature. Hence, to compare and illustrate the significance of introducing hydrogen to amorphous thin film V$_{80}$Zr$_{20}$, we justify the mentioned drift compensation. 
	
	\subsection{Theoretical calculations}
	Theoretical amorphous structures were created via the stochastic quenching method~\cite{Holmstr}, computed using \emph{ab initio} density functional theory (DFT) ~\cite{10.1103/physrev.140.a1133, 10.1103/physrev.136.b864}, as implemented in the Vienna Ab initio Simulation Package (VASP)~\cite{10.1103/physrevb.54.11169, 10.1016/0927-0256(96)00008-0, 10.1103/physrevb.47.558} with the generalized gradient approximation (GGA) correlation-functional of Perdew, Burke and Ernzerhof~\cite{10.1103/physrevlett.77.3865}.
	The stochastic quenching method can be used to construct maximally amorphous supercells by relaxing an initial configuration of randomized atomic positions, separated by at least one atomic radius, until the Hellman-Feynman forces acting on the atoms are below a predefined threshold, which was set to 20 meV/\AA~in the current study. Optimized structures are obtained by adjusting the system size of the supercells, allowing the atoms to be relax further, until the external stress is also below a threshold, which here was 1 kbar. To make sure that the system size fulfilled the non-self interaction criteria of Holmstr{\"o}m and co-workers~\cite{Holmstr}, a 14~\AA$~\times~$14~\AA$~\times$~14~\AA~supercell with 140 vanadium atoms and 35 zirconium atoms was constructed. Hydrogen of selected concentrations, $c_{\mathrm{H}} = \{0.2, 0.4, 0.6, 0.8, 1.0, 1.2, 1.4\}$ H/M, were introduced to the metallic glass structure while only allowing one of the lattice vectors of the supercell to expand freely, keeping the remaining lattice vectors fixed as to mimic the clamping behaviour present in the synthesized thin films. For each of these calculations, an energy cut-off of 450 eV with the Methfessel-Paxton smearing method~\cite{Paxton} of 0.2 eV width and a $\Gamma$-point-only $k$ sampling were used. Further details regarding the conditions, considerations and convergence of the volume expansion calculations can be found in Ref.~\cite{PhysRevB.106.104110}. 
	
	Once the set of hydrogen-loaded candidate structures had been created, they were each simulated at 300 K via \emph{ab initio} molecular dynamics (MD) simulations for at least $6.5$ ps to indirectly capture the phonon-electron coupling, a large contributor to the resistivity of metals~\cite{J_Jackle_1980}. The molecular dynamics simulations were carried out in the canonical NVT ensemble simulation scheme implemented in VASP, using the Nosé-Hoover thermostat with a Nosé mass of $0.4$. The effective mass was set initially to mimic the ionic movements typically found in vanadium, but was found to also yield stable temperature fluctuations and non-drifting ionic movements in these amorphous systems as well. Due to the computational cost, the energy cut-off and energy convergence criterion was set to $350$ eV and $5\cdot 10^{-4}$ eV throughout the MD calculations. Once the simulations had passed $2$ ps and the temperature fluctuations were stable, intermediate snapshot structures were extracted every $250$ fs, estimated to be sufficiently uncorrelated via the Bose-Einstein distribution weighted mean occupied frequency of the phonon density of states computed in Ref.~\cite{KAPLAN2022101496}. 
	
	The electronic total and orbital projected density of states (DOS) and frequency-dependent optical conductivity $\sigma(\omega)$, were also calculated using the implemented software in VASP. In total, $18$ thermalized snapshot structures per hydrogen-loaded structure were used for an averaged DOS and $\bm{\sigma}(\omega)$. For the DOS calculations, a $2\times2\times2$ $k$-points mesh with 350 eV as energy-cut-off and energy convergence criterion of $5\cdot 10^{-7}$ eV, with Fermi smearing width of $25.86$ meV was found to be sufficient for accurate convergence of the DOS. 
	The $s$-$p$-$d$-projected partial DOS were averaged across snapshots and categorized according to the atomic species and averaged across the total number of snapshots per hydrogen-loaded structure and put on a common scale by normalizing it with the volume of the structure. 
	
	The matrix elements of the frequency-dependent dielectric function, $\bm{\varepsilon}(\omega)$, were calculated in VASP via the random phase approximation. From a self-consistent field calculation using the same conditions as for the DOS calculations, the frequency-dependent dielectric function was calculated using $2.7$ times the default number of bands as to achieve sufficient convergence, and four different electron lifetime energy shifts ($\eta = \{0.02, 0.08, 0.16, 0.5\}$ eV) were considered to examine the indirect impact of excitation lifetime on the electronic properties. Once the matrix elements of $\bm{\varepsilon}$ had been calculated for each of the $18$ considered snapshots, the results were averaged. 
	
	The matrix elements of the frequency-dependent optical conductivity $\sigma_{\alpha\beta}$ can be computed via the relation\cite{alma991011857389707596, 10.3389/fphy.2022.946515}, 
	
	\begin{equation}\label{eq:Dielec_conduct}
		\varepsilon_{\alpha\beta} = \delta_{\alpha\beta} + i\frac{\sigma_{\alpha\beta}}{\varepsilon_0 \omega},
	\end{equation}
	
	\noindent
	where $\varepsilon_0$ is the vacuum permittivity. Evaluating the conductivity as a function of photon energies and taking the averaged trace of the reciprocal of the diagonal terms of $\sigma_{\alpha \alpha}(E)$ yield the energy-dependent resistivity, i.e. $\frac{1}{3}\Tr( \rho_{\alpha \alpha}(E))$. The DC resistivity was extrapolated down to $E = 0$ eV by a linear fit between $0.2 $ eV $< E < 0.6$ eV with two considerations in mind. One, the current-current and density-density  response functions should \emph{a priori} be mutually identical~\cite{PhysRevB.95.155203, giuliani_vignale_2005}, and we found that the two types of functions started to deviate below $0.2$ eV in the calculations where $\eta = 0.02$ eV. Secondly, the upper bound was chosen based on the condition that the considered region had to be sufficiently linear to yield within the standard deviation the same extrapolated $\rho_{\text{DC}}$. 
	
	The complex refractive index can be computed by taking the square root of the diagonal terms of $\varepsilon_{\alpha \alpha}$ according to~\cite{3056080},
	
	\begin{equation}\label{eq:Refract}
		N_{\alpha \alpha} = \sqrt{\varepsilon_{\alpha \alpha}} = n_{\alpha \alpha} + ik_{\alpha \alpha},
	\end{equation}
	
	\noindent
	where $n$ is the real refractive index and $k$ the optical extinction coefficient. The complex refractive index parallel to the expansion direction, i.e., $N_{z z}$, was used. 
	
	The theoretical transmission changes can now be constructed by using the right-hand side of Eq.~\ref{eq:meas_delta_k}, i.e., $k( c_{\text{H}} )t( c_{\text{H}} )/t_0 - k_0$. Highest quantitative agreement with the experimental results was found with a slightly blue-shifted extinction coefficient (by about 13.7~\%), along with the clamped experimental thickness expansion from Ref.~\cite{PhysRevB.106.104110}. Furthermore, the lifetime that gave best agreement corresponded to $\eta = 0.5$ eV, which is close to the experimental value for vanadium~\cite{PhysRevB.105.195438, FSacchetti_1983, PhysRevB.17.455}. The results are in Fig.~\ref{fig:Tr_vs_C}.

	
	
	\section{Results and Discussion} \label{sec:results}
	
	\subsection{Density of states}
\begin{center}
		\begin{figure}[t!] 
			\centering
			\includegraphics[width= 8.6cm]{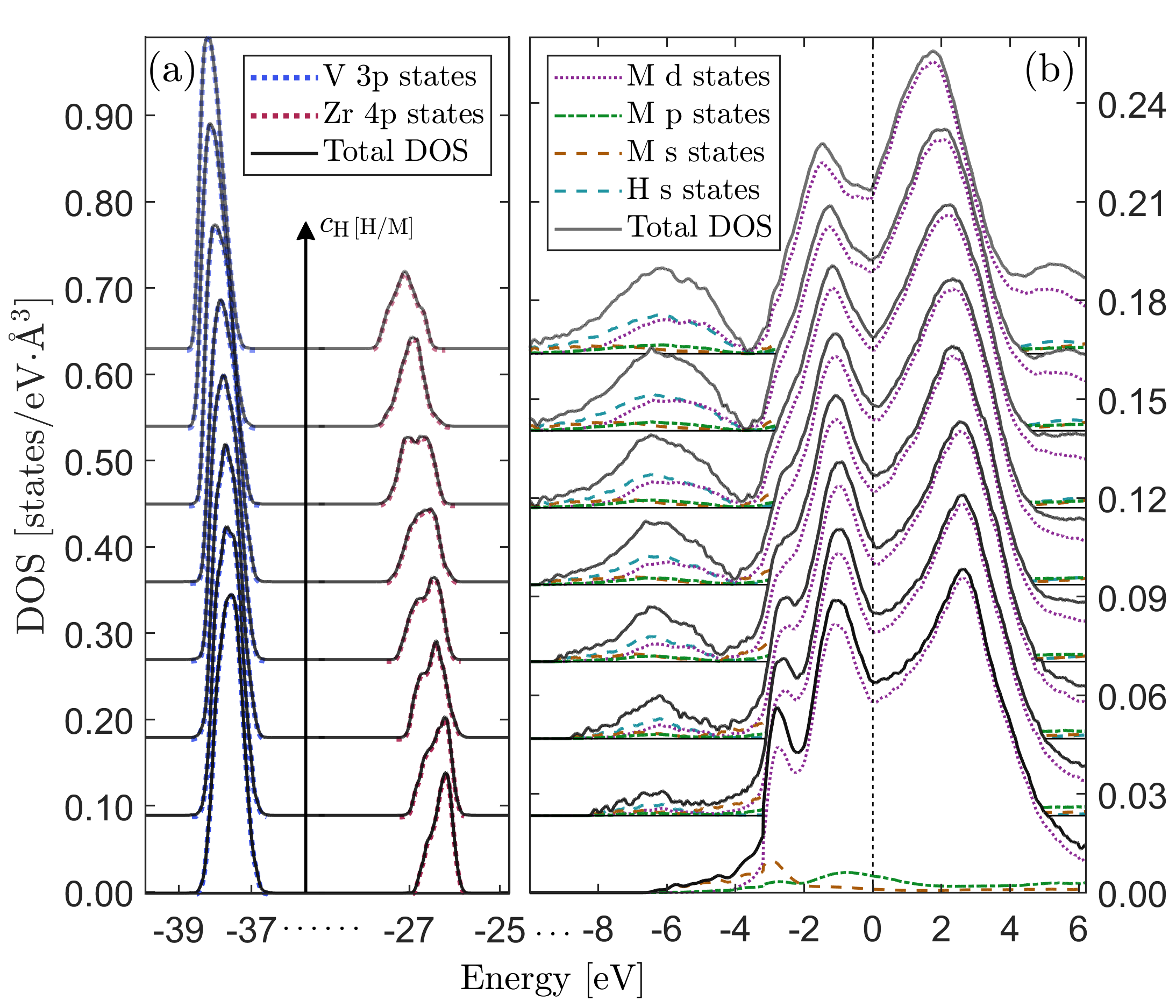}
			\caption{Theoretical DOS of V$_{80}$Zr$_{20}$, containing $c_{\mathrm{H}} = \{0.0, 0.2, 0.4, 0.6, 0.8, 1.0, 1.2, 1.4\}$ H/M concentrations of hydrogen. The $3p$ and $4p$ semicore states of vanadium and zirconium respectively are shown in (a), while (b) displays the total and $s$-$p$-$d$-decomposed metal M and hydrogen H DOS closer to the Fermi level.}
			\label{fig:DOS}
		\end{figure}
	\end{center}

	From our theoretical calculations, the total and partial $s$-$p$-$d$-decomposed DOS are shown in Fig.~\ref{fig:DOS} of the clamped hydrogen-induced expanded structures. Panel (a) of the figure includes the changes to the semicore $p$ states of the vanadium ($3p$) and zirconium ($4p$) for different concentrations of hydrogen, while panel (b) captures the changes in the valence band. In the $p$ semicore levels, there is an apparent shift with respect to hydrogen content, shifting from $-37.60$ eV to $-38.10$ eV for vanadium and $-26.27$ eV to $-27.10$ eV for zirconium. A seemingly linear scaling with hydrogen concentration shows up for the relative displacement of the zirconium states, while for vanadium, a non-linear trend is observed whereby the rate of the peak shift seems to increase at higher concentrations. By measuring the relative shift of the peaks and comparing the displacement with a known reference, it might be possible to deduce the hydrogen concentration in the metal. This could open up novel ways of measuring the hydrogen-induced properties of metal-hydrogen systems using photoemission techniques with gas loading capabilities. Furthermore, in Fig.~\ref{fig:DOS}(b), one can immediately see that the $d$ electrons dominate the pure metallic DOS, while both $s$ and $p$ states are comparably small. Some of the $d$ electrons from, in particular, the energy region around $-3$ eV, are lowered onto a level at around $-7$ eV with increasing hydrogen content. The $s$ electrons of the hydrogen atoms are also added to this part of the DOS, which shows that our \emph{ab initio} calculations predict that the $s$-$d$ hybridization found in many crystalline transition metal-hydrogen systems to be present in hydrogen-absorbed disordered V$_{80}$Zr$_{20}$ metallic glass as well. This suggests that the $s$-$d$ hybridization is not very sensitive to the local order of the atoms or the crystal symmetry of the interstitial sites. Interestingly, there are no occupied V $4p$ or Zr $5p$ states present in the outermost conduction states in the atomic electron configuration of either vanadium or zirconium, while here in the alloy they are present. This is also seen in crystalline hydrides of Ref.~\cite{PRXEnergy.3.013003}.
	
\begin{center}
		\begin{figure}[t!] 
			\centering
			\includegraphics[width= 8.3cm]{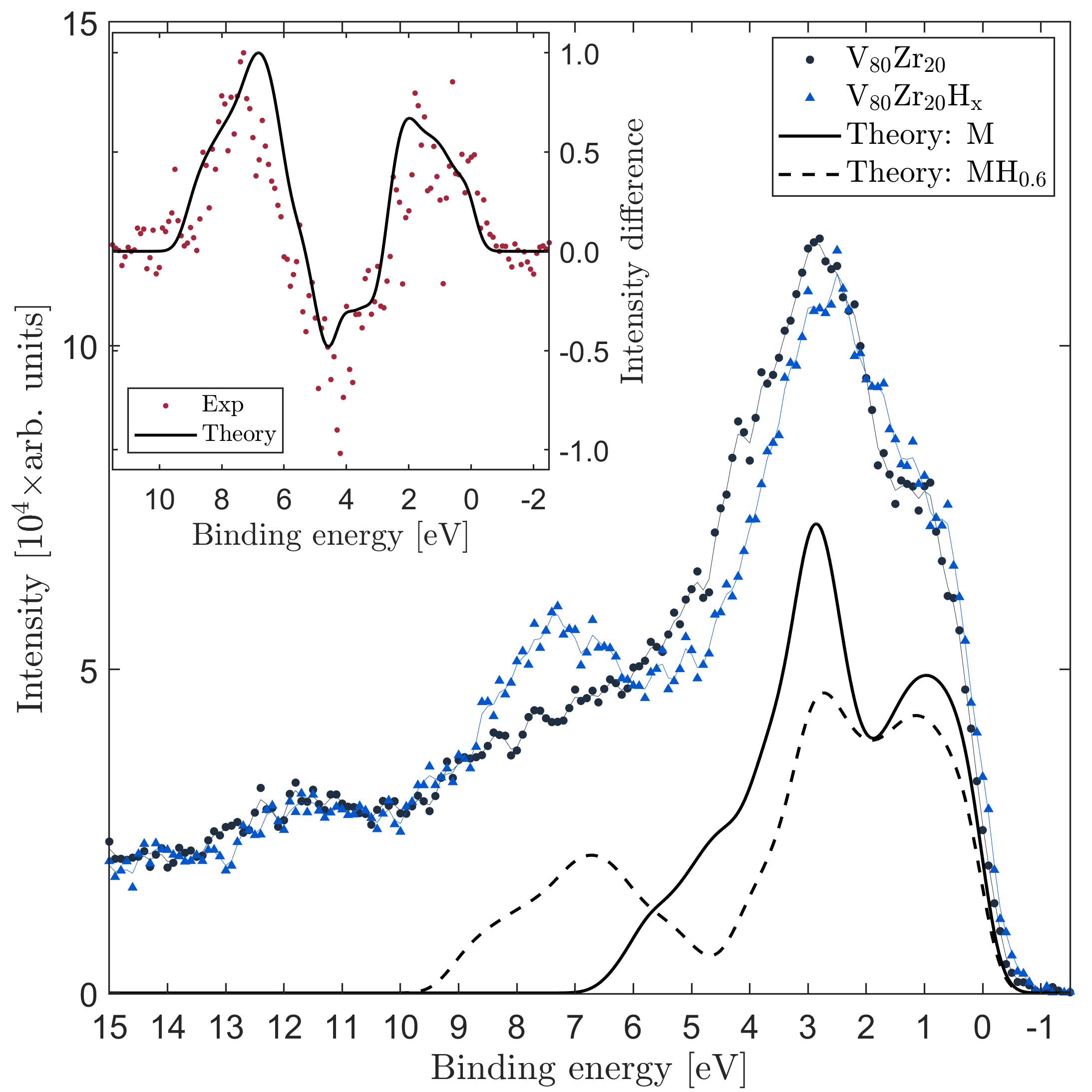}
			\caption{HAXPES intensities of V$_{80}$Zr$_{20}$, shown here as black circles, along with V$_{80}$Zr$_{20}$ loaded with hydrogen, represented here as blue triangles. The predicted theoretical cross-section-weighted DOS of the metal is shown in solid black line, while the black dashed line represents the weighted DOS of V$_{80}$Zr$_{20}$ with $0.6$ H/M of hydrogen.  The inset displays the relative change of the experimental curves together with the relative change of the theoretical curves.  The theoretical results have been scaled by a constant for ease of comparison.}
			\label{fig:HAXPES}
		\end{figure}
	\end{center}

	The HAXPES results using Ga-K$\alpha_1$ radiation are shown in Fig.~\ref{fig:HAXPES}. The intensity corresponding to the sample containing hydrogen is normalized to the background level of the scan of the sample without hydrogen under the assumption that background features are not significantly altered by hydrogen in the otherwise identically grown samples. Together with the experimental curves are the theoretical matrix element weighted total density of states for hydrogen-free  V$_{80}$Zr$_{20}$ and loaded V$_{80}$Zr$_{20}$H$_{0.6}$. The experimental resolution was determined to be $0.62$ eV and the theoretical curves, besides the intrinsic broadening of the calculated curves owing to the disordered nature of the structure, have been convoluted with a Gaussian of width $0.62$ eV as to mimic the instrumental smearing. The $s$-$p$-$d$ cross-sections of vanadium and zirconium evaluated at the Ga-K$\alpha_1$ energy were taken from the interpolated tables of Scofield~\cite{osti_4545040} as implemented in the Galore software~\cite{Jackson2018}, and while the presence of valence $p$ states is predicted in the theoretical DOS, there are no tabulated valence $p$ cross-sections for either vanadium or zirconium in their respective atomic configuration. This problem has also been encountered in a study on crystalline hydrides~\cite{PRXEnergy.3.013003} in which the $p$ cross-section was resolved via a heuristic approximation by assuming the cross-section being half of the valence $s$ cross-section, which we also have adopted here. Both theory and experiment predict an onset of spectral weight between 5 eV to 7 eV in binding energy with the addition of hydrogen. There are also subtle changes to the states between 0 eV and $5$ eV in binding energy that can be examined more closely. 
	The inset of Fig.~\ref{fig:HAXPES} shows the difference between the scan of the hydrogen-loaded sample and the intensity of the reference metal scan. In the same inset figure, presented as the solid black line, is the difference between the theoretical curves in Fig.~\ref{fig:HAXPES}. Both the experimental and the theoretical absolute difference have been normalized to 1 to better illustrate the predicted and observed relative changes. We find that both the experimental and the theoretical results showcase similar behavior, in which states have been added to the hydrogen-loaded metal in two regions, between 0 eV to 3 eV and 5 eV to 9 eV, while conversely, we observe between 3 eV to 5 eV a reduction in the occupied states.
	\begin{figure}[t!] 
		\centering
		\includegraphics[width= 8.3cm]{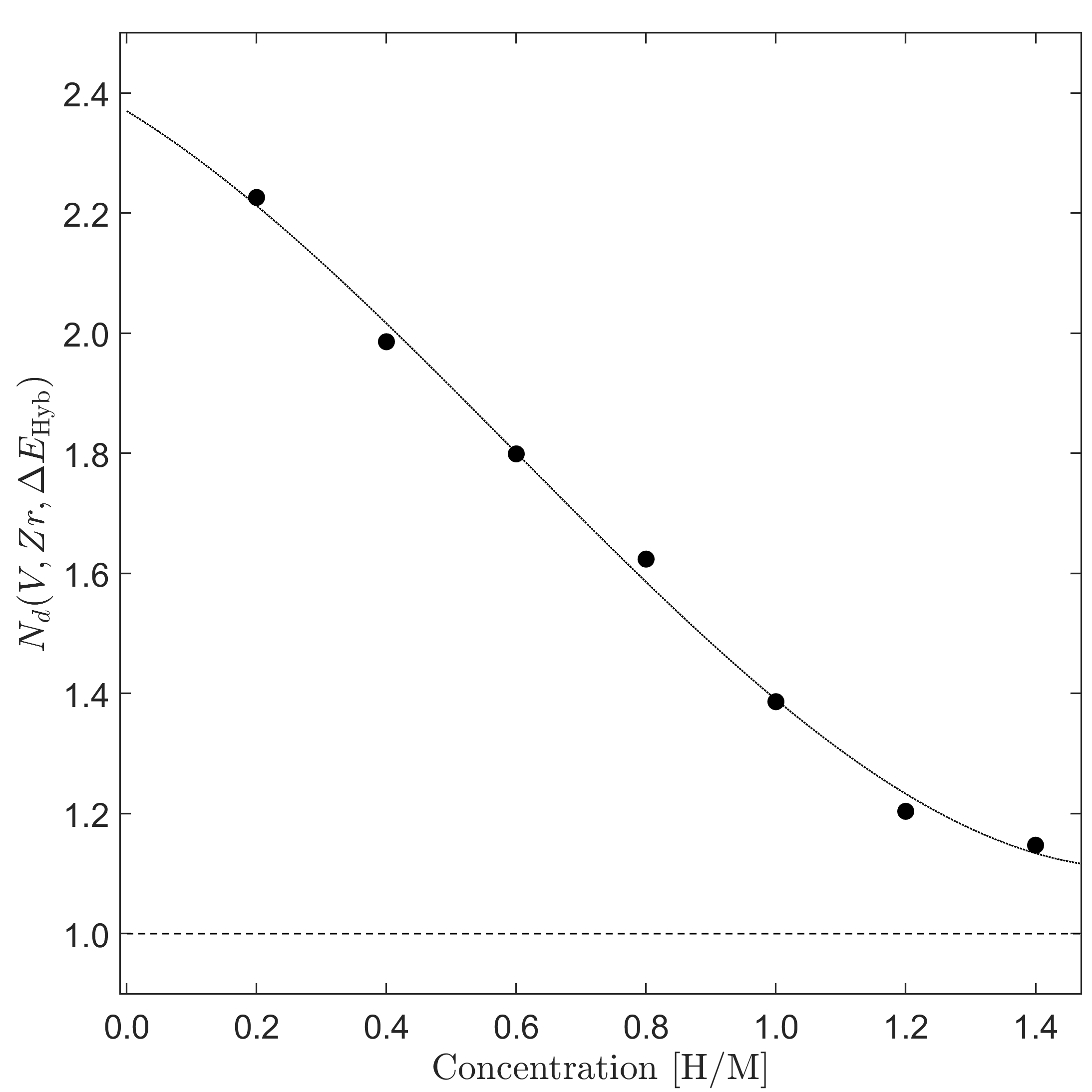}
		\caption{The ratio between the number of zirconium $d$ electrons and vanadium $d$ electrons participating in the hybridization around $E =$ $-$7$\pm$3 eV, computed via Eq.~\ref{eq:num_e}. Both the $d$ states of zirconium and vanadium has been normalized by their respective molar fraction composition. }
		\label{fig:Ratio_num_e}
	\end{figure}
	
	We can discern the relative elemental participation of the metal atoms in the $s$-$d$ hybridization with hydrogen by taking a ratio of the integrated number density of the vanadium and zirconium projected DOS of Fig.~\ref{fig:DOS} around  $E =$ $-$7$\pm$3 eV. For ease of discussion, we can therefore define the quantity:
	
	\begin{equation}
		N_{d}(i, j, \Delta E_{\mathrm{Hyb}}) = \frac{\bigintss\limits_{E \in \Delta E_{\mathrm{Hyb}}} DOS_{j, d}(E) dE }{\bigintss\limits_{E \in \Delta E_{\mathrm{Hyb}}} DOS_{i, d}(E) dE }~,
		\label{eq:num_e}
	\end{equation}
	
	\noindent
	where $d$ indicates that the quantity is with respect to the $d$ electrons, $i$ and $j$ represent the involved metallic species while $\Delta E_{\mathrm{Hyb}}$ represents the energy interval encompassing the hybridization of levels. After taking the vanadium and zirconium compositional differences into account, we show in Fig.~\ref{fig:Ratio_num_e} the relative participation in the hybridization between zirconium and vanadium.
	
	With a number density ratio greater than $2:1$ at low concentrations, we see that hydrogen has a clear preference for the zirconium local environment compared with the environment offered by vanadium. However, at higher concentrations the ratio levels out and hydrogen appears to have an almost equal affinity for both metal elements. This is consistent with the larger enthalpy of solution in crystalline ZrH$_{x}$~\cite{lewis1996hydrogen} as compared to VH$_{x}$~\cite{doi:10.1021/j100723a033, PhysRevB.58.5230} at low concentrations. It would therefore follow, that if the hydrogen kinetics in this material are to be improved, one route could be to lower the concentration of Zr in the alloy, or alternatively introduce a third element that pre-binds to the zirconium 4d states.
	
	We also measured the V $3p$ and Zr $4p$ semicore states and found that the Zr $4p$ peak shifted from a binding energy of about $E = 27.16(5)$ eV to $E = 27.41(5)$ eV, while no shift was observed for the V $3p$ peak within the resolution of the measurement. The displacement of the Zr peak supports the predicted theoretical binding energy scaling, and highlights the possibility of simultaneous \emph{in situ} thermodynamics and electronic measurements of metal-hydrogen systems using photoelectron spectroscopy. We note that hard x-ray photoemission in this energy range is much more sensitive to $s$ states and $p$ states of transition metals, which is an advantage when we want to see the $s$-$d$ hybridization but a disadvantage if one is interested in measuring the distribution of $d$ states in the material. 
	
	The level of agreement between theory and experiment regarding the energy interval of the hybridization implies that \emph{ab initio} DFT can to a large extent capture the general electronic structure changes of the metallic glass with the introduction of hydrogen.

	
	\subsection{Optical transmission}
\begin{figure}[t!] 
		\centering
		\includegraphics[width= 8.3cm]{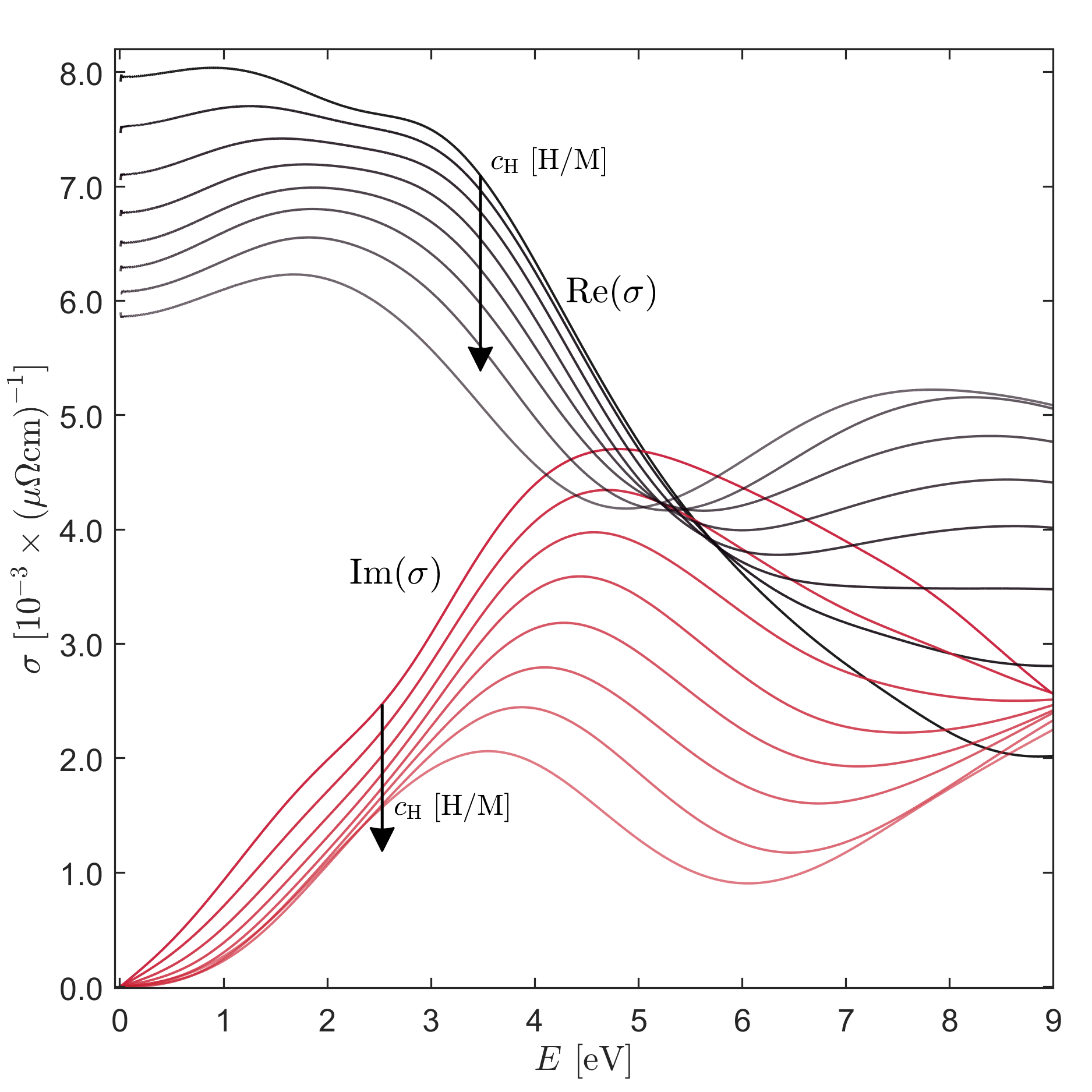}
		\caption{The real and imaginary component of the theoretical optical conductivity of V$_{80}$Zr$_{20}$, loaded with different concentrations of hydrogen. }
		\label{fig:conductivity}
	\end{figure}

	The combination of all allowed transitions between the occupied and the unoccupied electron states can be used to calculate the complex optical conductivity shown in Fig.~\ref{fig:conductivity}. Here, one can observe a general decrease in the conduction of electrons up to an energy around $5$ eV for different concentrations of hydrogen. However, around $E = 5.2$ eV, the trend flips and the material becomes more conductive with increasing contents of hydrogen. Another key feature that can directly be discerned from the figure is that the DC conductivity in the zero-energy limit decreases with hydrogen concentration. The change is most prominent for lower concentrations, but for higher concentrations, the relative increase in resistivity diminishes. 
	
\begin{figure}[t!] 
		\centering
		\includegraphics[width= 8.3cm]{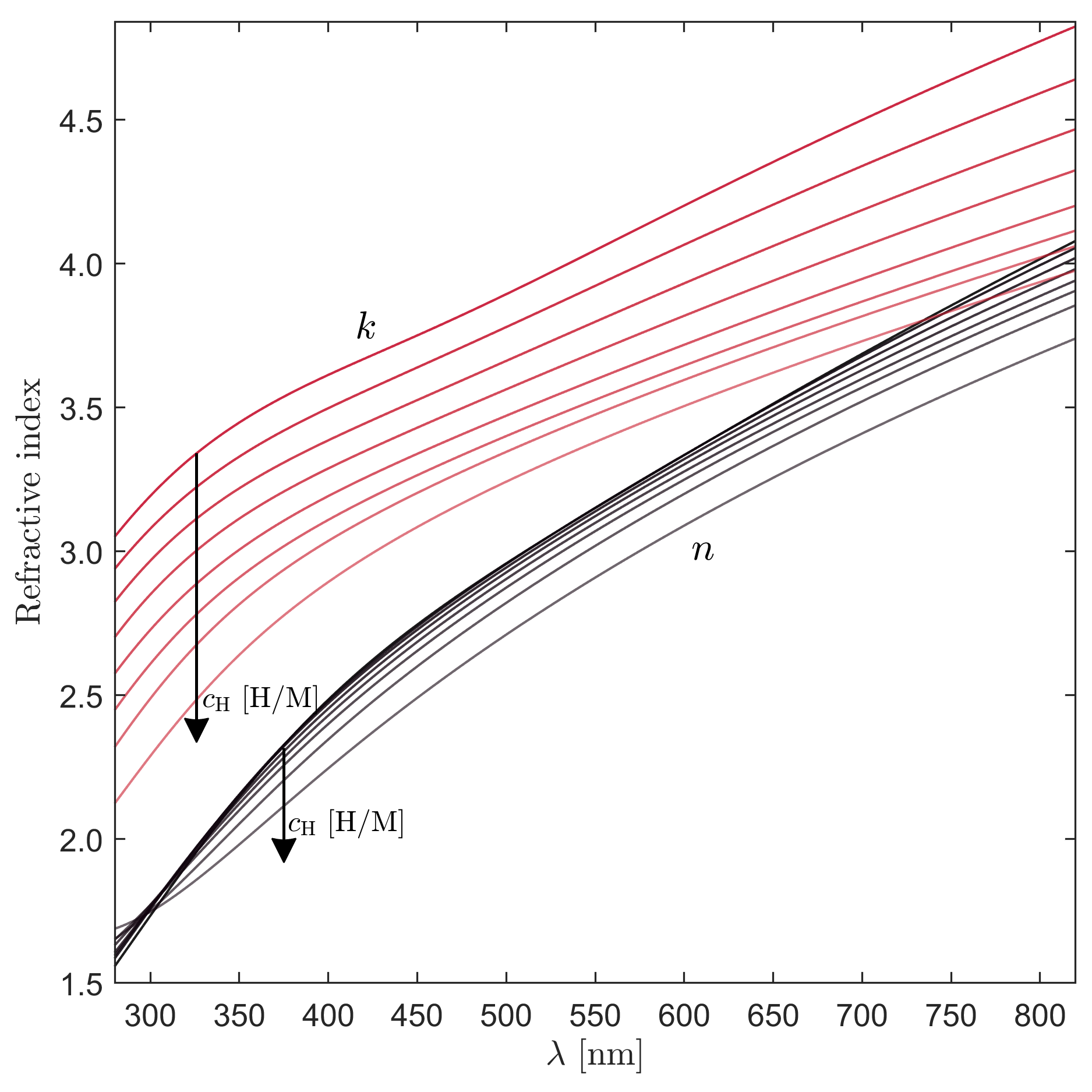}
		\caption{The components of the complex theoretical refractive index of V$_{80}$Zr$_{20}$ with different concentrations of hydrogen.}
		\label{fig:RefractIndex}
	\end{figure}

	The complex optical conductivity can be transformed into a corresponding complex refractive index using Eq.~\ref{eq:Dielec_conduct} and Eq.~\ref{eq:Refract}, which is shown in Fig.~\ref{fig:RefractIndex}. Except for the very highest hydrogen concentrations, the real part of the refractive index is to a large extent only weakly affected by the presence of hydrogen according to the results of the \emph{ab initio} calculations. However, a significant hydrogen dependence is instead observed in the imaginary part. The extinction coefficient $k$, predicts that with increasing concentration of hydrogen, the material absorbs fewer photons across the visible energy spectrum. It is worth noting here, that by tracing the relative changes, one can also observe a relative wavelength dependence of $k$ with respect to hydrogen concentration. In other words, the relative absorption decreases the most for photon wavelengths in the red and blue parts of the spectrum, while in the green range, the decrease is not as strong. The results presented here suggest that there should be an apparent wavelength dependence on the absorption coefficient when hydrogen is introduced to amorphous V$_{80}$Zr$_{20}$, which according to Eq.~\ref{eq:meas_delta_k}, should be reflected in the measured optical transmission. 
	
\begin{figure}[t!] 
		\includegraphics[width= 8.3cm]{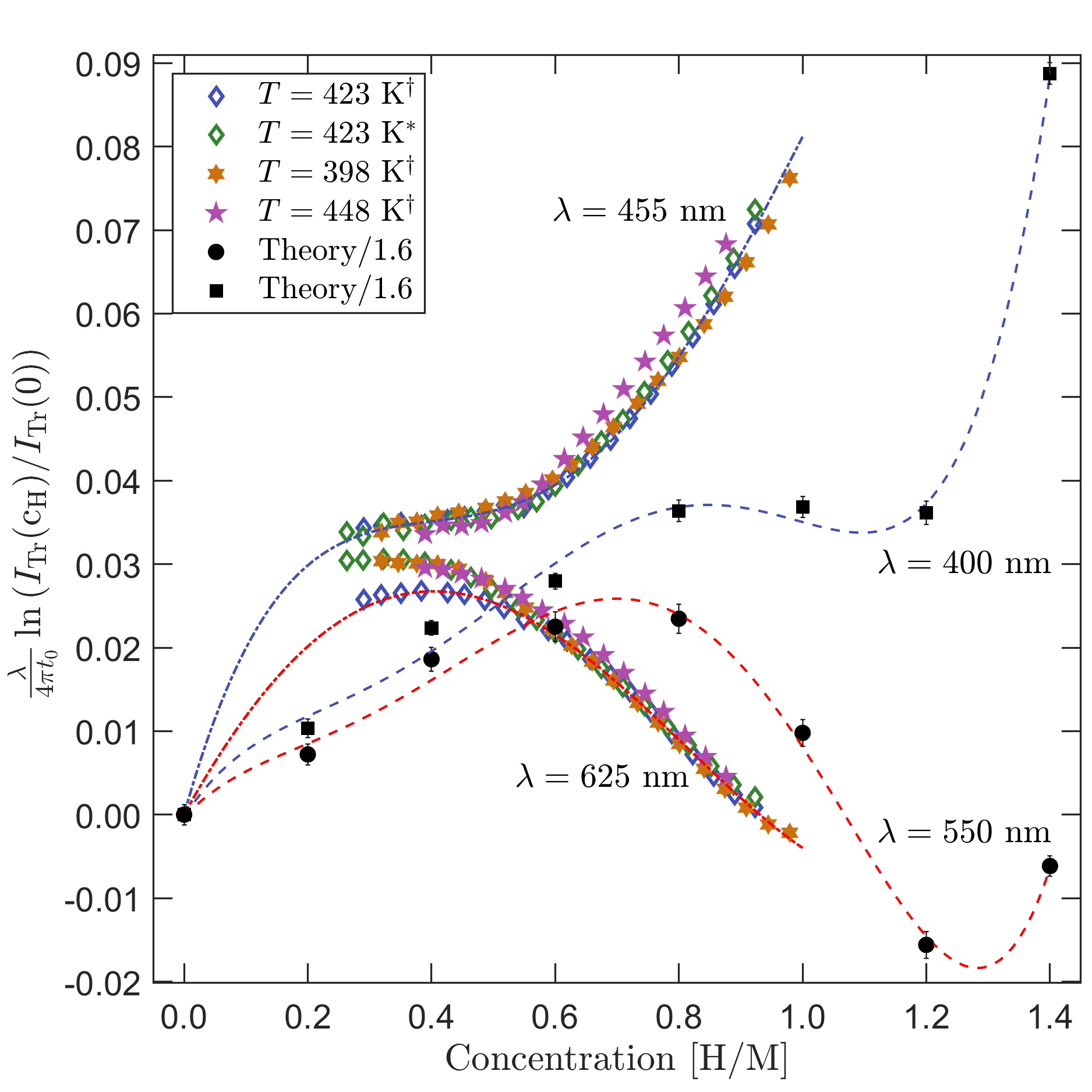}
		\caption{The changes in optical transmission, measured with a blue ($\lambda = 455$ nm) and red ($\lambda = 625$ nm) incoherent light source, at three different temperatures. The $\dagger$ superscript represents the first absorption cycle at a specific temperature while the $*$ superscript denotes the second cycle of that temperature. The subsequent cycles after the first $T = 423$ K one have been shifted to compensate for the time-dependent relaxation process. The corresponding theoretical predictions, using the experimental volume expansion~\cite{PhysRevB.106.104110} and blue-shifted by 13.7~\%, are displaced as the black markers, whereby their error bars represent the standard deviation originating from the MD averaging procedure.}
		\label{fig:Tr_vs_C}
	\end{figure}

	Summarized in Fig.~\ref{fig:Tr_vs_C} are the experimental optical transmission changes (at $\lambda= $ 625$\pm$25 nm and $\lambda=$ 455$\pm$25 nm) from four absorption-desorption cycles of hydrogen in a thin film of amorphous V$_{80}$Zr$_{20}$ presented in the form described by the left-hand side of Eq.~\ref{eq:meas_delta_k}, along with the theoretically predicted hydrogen-dependent transmission. Owing to the discrepancy between the experimental distribution of states as compared to the calculated ones in Fig.~\ref{fig:HAXPES}, we found that by blue-shifting the wavelength by about $13.7$ \% for both red and blue light, which is of the same order of magnitude as the spectral width of the diodes used, we could rationalize both theory and experiment. We still found it necessary to scale the calculated transmission changes by a factor of 1.6 to bring the calculated values closer to the experimental ones. This factor is not well understood at present but could be related to the used lifetime broadening value ($\eta = 0.5$ eV) being different from the actual experimental value. The deviation could also originate from limitations of the GGA functional and/or the DFT theory itself in describing the occupied and unoccupied states in the Kohn-Sham formalism. It is however unlikely that it is the disorder that is not  captured well by the model, as the stochastically quenched structures in Ref.~\cite{bylin2024oneshot} predict essentially identical pair distribution functions as measured ones. Furthermore, the quenched structures were also found in Ref.~\cite{PhysRevB.106.104110} to yield good agreement with the  observed hydrogen-induced global expansion of amorphous V$_{80}$Zr$_{20}$.

	\begin{figure}[t!] 
		\centering
		\includegraphics[width= 8.3cm]{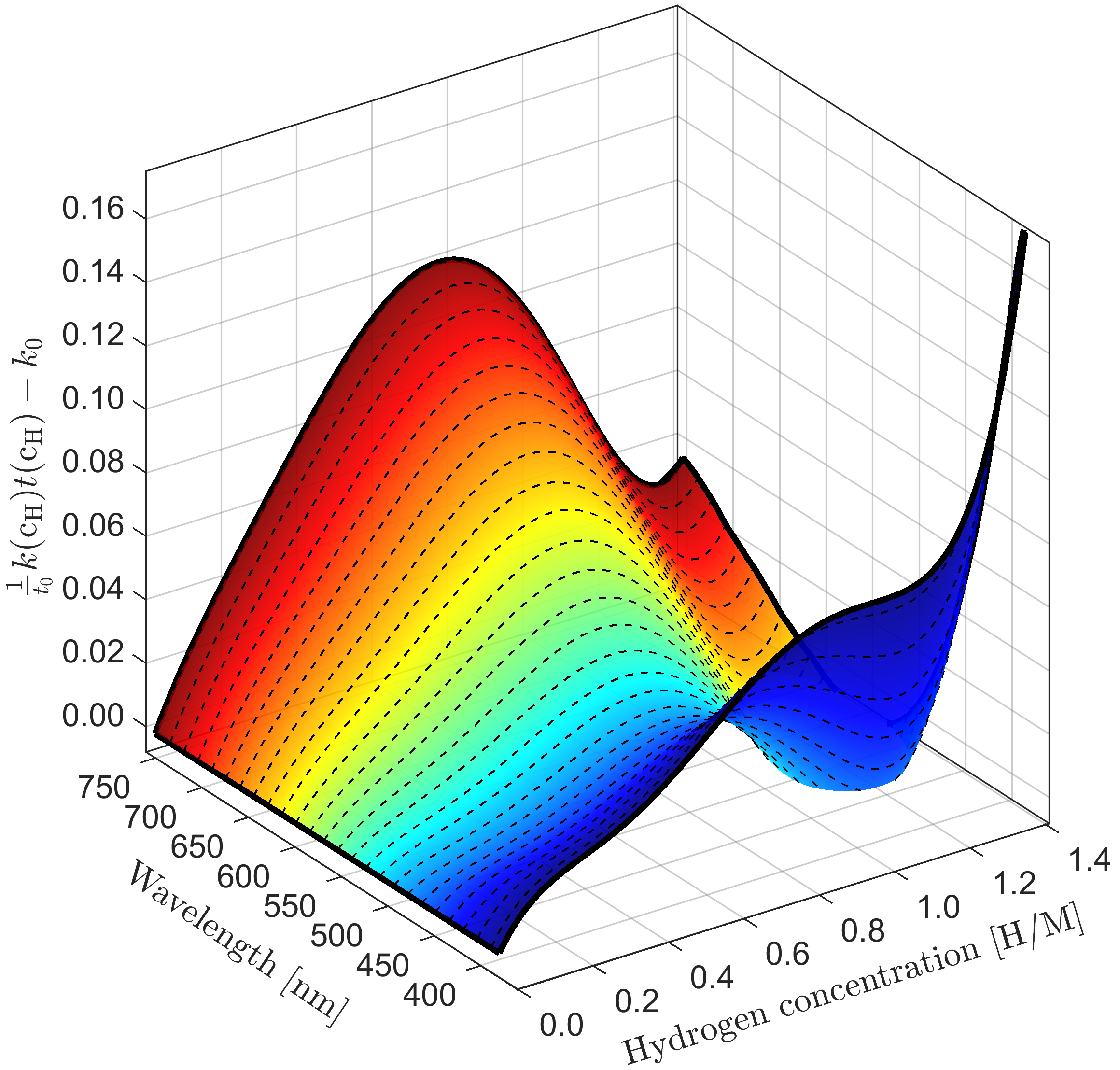}
		\caption{The theoretical optical transmission for a continuum of wavelengths. The experimental volume expansion from Ref.~\cite{PhysRevB.106.104110} has been implemented. }
		\label{fig:Delta_K_all}
	\end{figure}
	Nevertheless, one can observe in Fig.~\ref{fig:Tr_vs_C} a stark difference when it comes to the hydrogen-induced relative change between the measured red and blue light in transmission. The collected blue data exhibits a curve with features similar to that of a saddle point function. Contrary to the blue light transmission, the observed transmission in the red part of the optical spectrum forms a parabola whose maxima almost aligns where one observes the saddle point of the blue light transmission.
	
	Comparing the experimental results with the theoretically computed transmission changes we find qualitative agreement regarding the displayed parabolic and saddle point features. We note a concentration offset in the calculated curves, and this discrepancy suggests that the effect of hydrogen on the optical properties is underrepresented in the DFT calculations, as per the already mentioned possible reasons. A similar, and possibly linked conclusion, has been offered for the disparity between experimental and theoretical hydrogen dipole forces and volume expansions, as studied in Ref.~\cite{PhysRevB.106.104110, 10.1016/s1359-6454(02)00173-8, 10.1103/physrevb.96.224103, 10.1103/physrevb.94.241112}. This deficit could underestimate the shifts of the energy bands computed for these metal-hydrogen systems, which in turn translates into an indirect underrepresentation of hydrogen-induced changes to the optical conductivity.
	
	As predicted from the complex refractive index of Fig.~\ref{fig:RefractIndex}, the absorption coefficient should decrease irrespective of hydrogen concentration or choice of wavelength. However, once combined with the thickness expansion of the material, one observes that the decreasing absorption is compensated by the expansion to form effective parabolas around the red and green parts of the visible spectrum. This interplay between the varying extinction coefficients and volume expansion can further be scrutinized by examining how they combine into a continuous function of both hydrogen concentration and wavelengths. In Fig.~\ref{fig:Delta_K_all} we graph these apparent transmission changes. Here, as a note of clarification, the slices of curves along $\lambda = 400$ nm and $\lambda = 550$ nm are the ones displayed for direct comparisons with experimental data in Fig.~\ref{fig:Tr_vs_C}, although it becomes evident in this continuous graph that the general features observed experimentally are qualitatively captured at slightly more blue-shifted wavelengths. One can immediately conclude that there is a significant wavelength dependence to the observed optical transmission. This is in stark contrast to what has been observed in crystalline vanadium, whose transmission changes have been shown to decrease with hydrogen concentration, and to be largely wavelength-independent and mainly driven by the volume expansion alone~\cite{10.1063/1.3529465}. This highlights that one can easily not infer the optical properties of metallic glass by simply inheriting the features found in its closest crystalline counterpart.

	\subsection{DC conductivity}
	
\begin{figure}[t!] 
		\centering
		\includegraphics[width= 8.3cm]{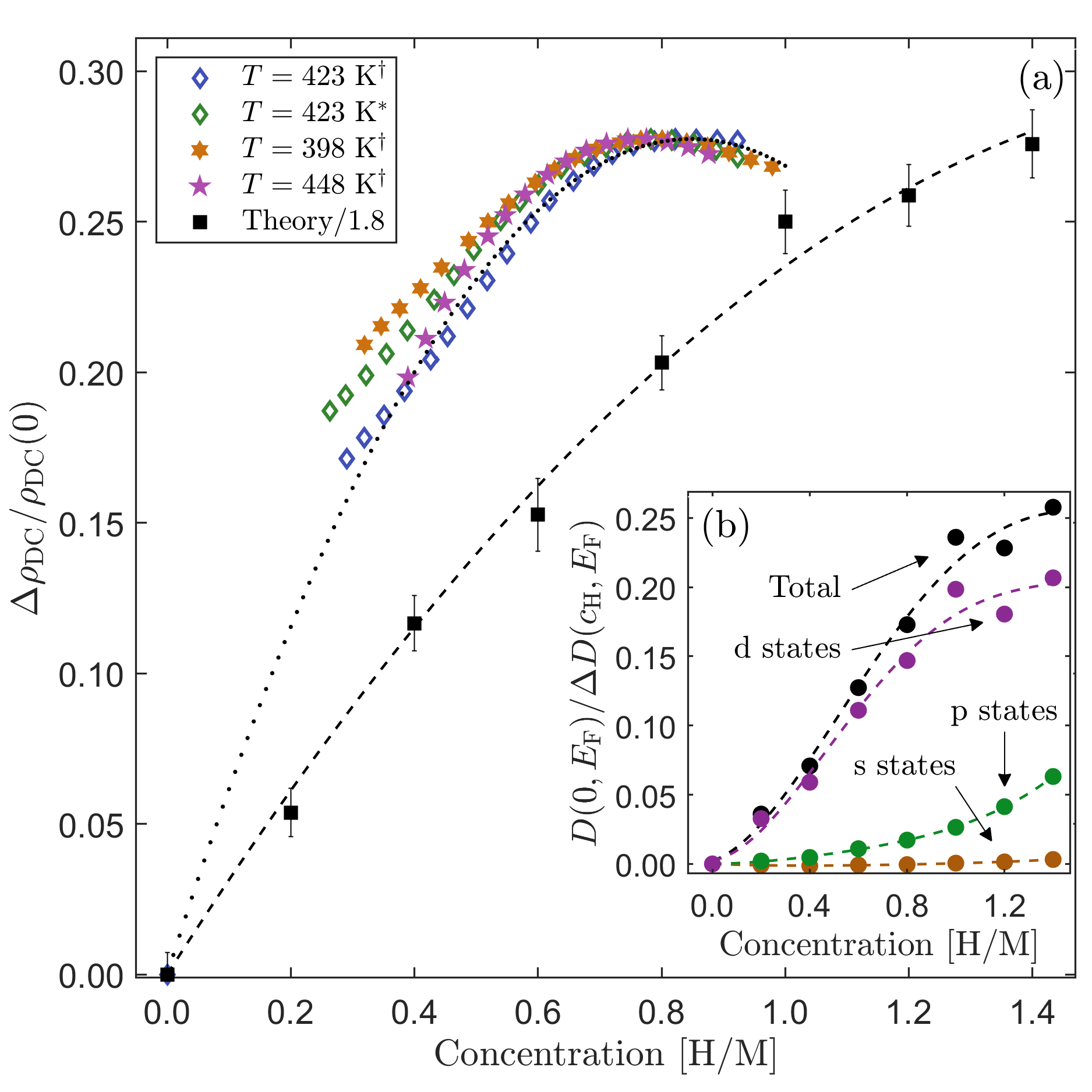}
		\caption{(a) Changes in resistivity of V$_{80}$Zr$_{20}$ for different concentrations of hydrogen, measured at three different temperatures. The $\dagger$ superscript represents the first absorption cycle at a specific temperature while the $*$ superscript denotes the second cycle of that temperature. The subsequent cycles after the first $T = 423$ K one have been shifted to compensate for the time-dependent relaxation process. The corresponding theoretical prediction is displayed as black square markers, and the error bars represent the standard deviation derived from the MD averaging procedure. The dotted and dashed lines are the fits to the model by Pryde and Tsong~\cite{PRYDE19711333}. \\
			(b) The hydrogen-induced changes in the reciprocal of the total and $s$-$p$-$d$-decomposed DOS evaluated at the Fermi level.}
		\label{fig:R_vs_C}
	\end{figure}

	Turning the attention to the hydrogen-induced changes to the resistivity, shown in Fig.~\ref{fig:R_vs_C}(a), we observe that the resistivity of both the experimental measurements and the theoretical calculations increases with hydrogen concentration, which has been recorded in other metal-hydrogen systems~\cite{KOKANOVIC1999795, 10.1063/1.3529465}. The resistivity appears to follow a parabola with a maximum appearing around $0.8$ H/M. This is similar to the behavior of resistivity in the transition metal hydrides such as vanadium and tantalum in the disordered $\alpha$ phase in the concentration range from $0.0$ H/M to $0.7$ H/M~\cite{PRYDE19711333}. Also, the theoretical resistivity curve, besides being scaled by a factor of $1.8$ for comparison with the experiment, showcases a similar concentration offset as the one that was found in the optical transmission. As both the computed refractive index and DC resistivity originate from the same optical conductivity tensor, it is not surprising that the changes in resistivity are also subject to a similar underrepresentation as found with the optical transmission. However, there could also be other unaccounted mechanisms behind the observed discrepancy than those already mentioned~\cite{PhysRevB.105.104114}, such as electron scattering phenomena~\cite{NAUGLE1984367, https://doi.org/10.1002/pssa.2210170217} and possibly weak localization~\cite{dugdale_1995, PhysRevB.41.958} that the current formalism fails to capture. We can analyze the resistivity changes in terms of the number of available sites for hydrogen by fitting the resistivity using the simple model by Pryde and Tsong~\cite{PRYDE19711333}, for which we find that the model suggests a maximum filling of about $1.7$ H/M for the experimental results and $3.7$ H/M for the theoretical curve. While a maximum of $1.7$ H/M does not break past the $2.0$ H/M barrier in transition metal hydrides, for which there are almost no exceptions, in terms of the elemental embedding in a homogeneous medium, by virtue of effective medium approximations~\cite{choy2016effective, BANKMANN2003566, PhysRevB.38.3690}, a higher concentration of zirconium whose negative enthalpy of hydrogen solution~\cite{lewis1996hydrogen} could possibly drive the maximum hydrogen concentration of the amorphous vanadium-zirconium system past this $2.0$ H/M limit.
	
	According to the Drude formula, the decrease of conduction states around the Fermi level should reflect the calculated increase in resistivity. Evaluating the DOS at the Fermi level for each hydrogen-loaded structure and taking the relative reciprocal change, i.e. $D(0,E_{\text{F}})/\Delta D(c_{\text{H}},E_{\text{F}})$, we observe in  Fig.~\ref{fig:R_vs_C}(b) that the diminishing of states does indeed inhibit the conduction of electrons and is essentially proportional to the resistivity change calculated in Fig.~\ref{fig:R_vs_C}(a). This further strengthens the argument that the noted electronic underrepresentations with respect to hydrogen concentration originate from a deeper level of DFT computation complexity.
	
	
	\section{Conclusions} \label{sec:conclusion}
	The hydrogen-induced $s$-$d$ hybridization observed in most crystalline transition metal hydrides persists in glassy V$_{80}$Zr$_{20}$H$_x$ as seen from the HAXPES results as well as from the \emph{ab initio} calculations. The calculations confirm that the changes $-7$ eV below the Fermi level are indeed associated with both H $1s$ and V $3d$ and H$1s$$-$Zr$4d$ hybridization. This consistent presence of the spectral weights at approximately the same binding energy suggests that this feature is universal in transition metal-hydrogen systems and is not sensitive to global crystal symmetries, including crystal-field splittings. Joint \emph{in situ} thermodynamic and electronic studies of the amorphous metal-hydrogen systems are proposed to be possible with photoelectron spectroscopy by measuring and relating the shift in the metal core levels to an effective hydrogen concentration in the material. The accompanied decrease in the states at the Fermi level shown in Fig.~\ref{fig:DOS} and its connection to the resistivity displayed in Fig.~\ref{fig:R_vs_C} signifies that electronic transport could be explained in terms of the prevalence of $d$ states, which might make it seem as if the $d$ states are responsible for the conduction in this material. However, the decrease in the $d$ states could also be a sign of the indirect addition of hydrogen which subsequently hybridizes to decrease the $d$ states, when in fact the increase in resistivity could be caused by the scattering of electrons via the perturbation to the potential landscape induced by the hydrogen; these two possibilities cannot currently be distinguished by us within the scope of this article. 
	
	The volume expansion plays a significant role in how emergent electronic features present themselves in terms of macroscopic observables, such as resistivity or optical transmission changes. In crystalline vanadium, the observed changes in optical transmission are dominated by volume changes, while the modifications in the spectral weight of the occupied and unoccupied states have less impact on the optical transmission. In V$_{80}$Zr$_{20}$ on the other hand, we find that both effects are important, yielding a strong concentration dependence on the apparent transmission. 
	
	The experimental and theoretical results show that the method of using optical conductivity to determine hydrogen concentration in films~\cite{10.1063/1.2821376, 10.1063/1.3529465, https://doi.org/10.1002/adma.200602560} can  be fruitful in metallic glasses. This opens up the possibility of using optical techniques to screen also disordered materials for desirable hydrogen properties, such as rapid diffusion or thermodynamic properties. Furthermore, the theoretical work presented here provides a starting point for other types of calculations, such as the electronic interaction between hydrogen atoms~\cite{10.1103/physrevb.27.7575}, which can be used to predict the enthalpy of solution~\cite{GelattC.D.1975Hofo} and the entropy of the metal-hydrogen systems~\cite{10.1063/1.1410114}.

	
	\section*{Acknowledgements} \label{sec:acknowledgements}
	The authors gratefully acknowledge the Swedish Research Council (VR -  Vetenskapsrådet) grant number 2018-05200 for funding the work and the Swedish Energy Agency grant 2020-005212. G.K.P. also acknowledges the Carl-Tryggers foundation grants CTS 17:350 and CTS 19:272 for financial support. The computations were enabled by resources provided by the Swedish National Infrastructure for Computing (SNIC), partially funded by the Swedish Research Council through grant agreement no.\ 2018-05973.
	\bibliography{bib_PRM.bib}
	
\end{document}